\documentclass[aps,twocolumn,prl,showpacs]{revtex4}

\usepackage{amsbsy}
\usepackage{graphicx}
\begin{document}
\newcommand{\beq}{\begin{equation}}
\newcommand{\eeq}{\end{equation}}
\newcommand{\bear}{\begin{eqnarray}}
\newcommand{\eear}{\end{eqnarray}}
\renewcommand{\a}{\alpha}
\renewcommand{\b}{\beta}
\newcommand{\tb}{\tilde{\beta}}
\newcommand{\rd}{\mathrm{d}}
\newcommand{\lp}{\left}
\newcommand{\rp}{\right}
\newcommand{\bs}{\mathbf}
\newcommand{\bz}{\bs{\hat{z}}}
\newcommand{\bx}{\bs{\hat{x}}}
\newcommand{\by}{\bs{\hat{y}}}
\newcommand{\hb}{\hbar}
\newcommand{\g}{\gamma}
\newcommand{\pd}{\partial}
\renewcommand{\l}{\lambda}
\renewcommand{\d}{\delta}
\renewcommand{\o}{\omega}
\renewcommand{\O}{\Omega}
\newcommand{\ofm}{\o_{fm}}
\newcommand{\s}{\sigma}
\renewcommand{\bm}{\bs{M}}
\newcommand{\sm}{\bs{m}}
\newcommand{\bmo}{\bs{M_0}}
\newcommand{\bmp}{\bs{M_\perp}}
\newcommand{\smp}{\bs{m_\perp}}
\newcommand{\bmx}{\bs{M_x}}
\newcommand{\bmy}{\bs{M_y}}
\newcommand{\smx}{\bs{m_x}}
\newcommand{\smy}{\bs{m_y}}
\newcommand{\nb}{\nabla}
\newcommand{\tm}{\times}
\newcommand{\bb}{\bs{B}}
\renewcommand{\sb}{\bs{b}}
\newcommand{\bbz}{\bs{B_z}}
\newcommand{\ba}{\bs{A}}
\newcommand{\bk}{\bs{k}}
\newcommand{\bh}{\bs{H}}
\newcommand{\sh}{\bs{h}}
\newcommand{\z}{\zeta}
\newcommand{\mxo}{M_{x\,1}}
\newcommand{\mxd}{M_{x\,2}}
\newcommand{\mxt}{M_{x\,3}}
\newcommand{\myo}{M_{y\,1}}
\newcommand{\myd}{M_{y\,2}}
\newcommand{\myt}{M_{y\,3}}
\newcommand{\mxi}{M_{x\,i}}
\newcommand{\myi}{M_{y\,i}}
\newcommand{\mxy}{M_{x,y}}
\newcommand{\oth}{\o_{th}}
\newcommand{\Oth}{\O_{th}}
\newcommand{\lt}{\tilde \l}
\newcommand{\se}{\bs{e}}
\newcommand{\pp}{{(+)}}
\newcommand{\mm}{{(-)}}
\newcommand{\ppm}{{(\pm)}}
\newcommand{\mpm}{m^{(\pm)}}
\newcommand{\bpm}{b^{(\pm)}}
\newcommand{\hpm}{h^{(\pm)}}
\newcommand{\epm}{e^{(\pm)}}
\newcommand{\zpm}{\z^{(\pm)}}
\newcommand{\zr}{\z_{res}^{(+)}}
\newcommand{\im}{\mathrm{Im}}
\newcommand{\re}{\mathrm{Re}}
\newcommand{\pr}{Phys.~Rev.~}
\setlength\arraycolsep{1pt}

\title{Excitation of Spin Waves in Superconducting Ferromagnets }
\author{V.~Braude}
\author{E.~B.~Sonin}
\affiliation{The Racah Institute of Physics, The Hebrew University of
Jerusalem,
Jerusalem 91904, Israel}
\date{\today}
\begin{abstract}
This Letter presents a theoretical analysis of propagation of spin waves
in a superconducting ferromagnet.
The surface impedance was calculated  for the case when the
magnetization is
normal to the sample
surface. We found the frequencies at which the impedance and the
power absorption
have singularities related to the spin wave propagation,  and determined
the form
of these singularities. With a
suitable choice of parameters, there is a frequency interval in which two
propagating spin waves of the same circular
polarization are generated, one of them having  a negative group
velocity.
%, i.e., directed opposite
% to the phase velocity.
%The results of the theory demonstrate that
%the microwave probing of spin
%dynamics can be an effective experimental method to study materials with
%coexisting  superconducting
%and  ferromagnetic orders.
%, including those called superconductors with
%broken time-reversal symmetry.
\end{abstract}
%\begin{document}
%\title{ferromagnetic resonance}
\maketitle
%\section{Introduction}
Coexistence of superconductivity (SC) and ferromagnetism
(FM) has been a long-standing problem since the beginning of the modern
theory of SC~
\cite{ginz,anderson,BulBuz}. There has been a
renewed interest to this problem, and SC-FM coexistence was discovered recently
in various materials such as ruthenocuprates \cite{felner,bernhard},
$\mathrm{ZrZn}_2$ \cite{pfleiderer}, $\mathrm{UGe}_2$ \cite{saxena}, and
$\mathrm{URhGe}$ \cite{aoki}.
%and some others.
It was also suggested theoretically that in $p$-wave superconductors, time
reversal
symmetry is broken, and a non-zero
magnetic moment arises \cite{mac}.

The existence of  the FM order in a superconductor is hidden by Meissner
currents, which create a magnetic
moment opposite to the  spontaneous magnetic moment. Moreover, in the
Meissner state SC eliminates another consequence of the FM order: the
equilibrium domain structure \cite{DS}.  However,
the Meissner currents cannot
shade such an important manifestation of the FM order as spin waves. A
conventional technique to  probe spin modes in  ferromagnets, both
insulators and
metals, has been observation of Ferromagnetic Resonance (FMR) \cite{akhi}.
There
are also
reports on FMR observations in materials with SC--FM coexistence
\cite{fmr-sc}.

In this work we theoretically investigated propagation of spin waves
in a material, which possesses both SC and FM order and is  irradiated by
an electromagnetic (EM) wave. The
first step in this direction was done by Ng and Varma \cite{NV}, who
studied the spectrum of  spin
waves interacting with  vortex modes in the {\em spontaneous vortex
phase} (mixed state in zero
external magnetic field). Here we analyze the boundary
problem for the Meissner state: how  an incident EM wave  can
generate spin waves inside a sample. We solved the Landau-Lifshitz equation
for the
magnetization and the equations of London electrodynamics, assuming that
the equilibrium
magnetization is normal to the surface and there is no external
static magnetic field and, as a
result of it, no Meissner currents at equilibrium.
We found the spectrum
of  spin waves for two cases,  depending on the stiffness of the spin system.
If the stiffness is large
enough, the spectrum
looks similar to that of a FM insulator, with the minimum wave
frequency (threshold for spin wave
propagation) equal to the frequency of the uniform FMR,
i.e. corresponding to zero
wave vector \cite{akhi}. But the opposite case of small spin stiffness is  more
interesting. Then  propagation of spin waves becomes possible at
frequencies lower than
that of the uniform FMR, and  waves near the threshold have finite wave
vectors.
Moreover, a monochromatic  incident EM wave generates {\em
two} propagating spin waves with
slightly different wave vectors. One of them has a negative group
velocity, i.e., its direction is
opposite to that of the phase velocity.  We found the
surface impedance, which is intimately connected with the spin wave
spectrum. The analysis was focused on the
case without explicit dissipation terms in the dynamical equations, when
 all energy absorbed from the incident radiation is
%absorption of the energy  is exclusively
%due to the energy
carried  away by spin waves propagating deep into the
sample bulk.
But we discuss also the role of various dissipation sources. The analysis
demonstrates
that the microwave probing of spin modes, which has been so fruitful
for normal ferromagnets,  should be also an effective technique for
investigation of
unusual properties of spin modes in
superconducting ferromagnets (SCFM's).

\begin{figure}
\begin{center}
\includegraphics[width=0.3\textwidth]{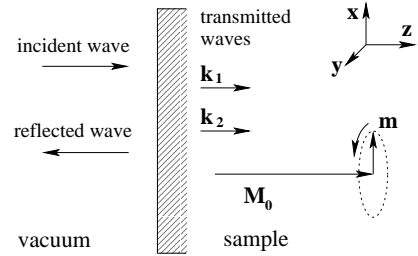}
\caption{The setup geometry.}
\label{setup}
\end{center}
\end{figure}

We consider a SCFM occupying the
semispace $z>0$ as shown in Fig.~\ref{setup}. This is a good
approximation for a
slab so thick that one can neglect a possible reflection of waves
from the second
boundary.
%Hence the slab can be replaced by a semispace, and our geometry is
%shown in Fig.\ref{setup}.
Uniting  the micromagnetism approach for ferromagnets and the London
approximation for
superconductors leads to the following free energy functional for our  
material:
\bear
  F&=&\int \rd^3x \Bigg\{2\pi \a \left[M_{\perp}^2+l_d^2(\pd_i
M_j)(\pd_i M_j)\right]
  \nonumber \\ && \qquad
 + \frac{1}{8 \pi \l^2}\lp(
 \frac{\Phi_0}{2\pi}\mathbf {\nabla} \phi-\bs{A}\rp)^2+\frac{B^2}{8\pi}
-\bs{B\cdot M}
%-\frac{\bs{B\cdot H}}{4\pi}
\Bigg\}~,
\eear
where $\phi$ is the phase of the SC
order parameter, $\Phi_0=h c/2 e$ is the magnetic flux quantum,  $\bm$ is
the FM
magnetization, and $M_\perp$ is the magnetization
component perpendicular to the easy axis $\bz$, which is normal to the
sample surface.
At equilibrium the magnetization is
$\bs{M_0}=M_0
\,\bz$ ($M_\perp=0$), and the large anisotropy parameter
$\alpha >1$ (the ratio
of the anisotropy energy to the magnetostatic energy) ensures that our
configuration is stable against
the flip
of $\bmo$ into the plane parallel to the surface.
 In our geometry the static magnetic
induction $\bs{B_0}$ vanishes both inside and outside the sample, but the
magnetization creates a
magnetic field inside the sample: $\bs{H_0}=-4\pi \bmo$. In
accordance
with the micromagnetism approach,
the absolute value of
$\bm$ is assumed to be constant $M=M_0$, and hence terms dependent on it were
omitted.  The length $l_d$,
which is of the order of the domain wall thickness in normal ferromagnets,
characterizes the
stiffness of the spin
system. We consider a single-domain sample, since in SCFM's  in the Meissner
state there are no
domains at equilibrium  \cite{DS}.

The spin dynamics is governed by the Landau-Lifshitz equation \cite{akhi}
\beq
  \frac{\rd \bs{M}}{\rd t}=-g\lp(\bs{M}\times \frac{\d F}{\d \bs{M}} \rp).
 \label{LL}\eeq
where $g$ is the gyromagnetic factor. Since we are concerned here with motion
near the equilibrium, we can decompose
\beq
  \bm=\bmo+ \sm,
\eeq
where
$\sm \perp \bmo$ is the dynamic part.
For plane waves $ \propto e^{ikz-i\omega t}$, Eq.
(\ref{LL}) yields:
\beq \label{eq:dm}
-i\omega  \sm=-g \bmo \tm \lp\{4\pi\a (1+l_d^2 k^2)\sm- \sb\rp\},
\eeq
where, since  the
equilibrium magnetic induction vanishes, there is only a dynamic part of
the induction: $\mathbf B=\sb$.
To find $\sb$, we minimize the free energy with respect to the vector
potential $\ba$.
This yields the London equation:
\beq \label{eq:london}
   \lp(1+k^2 \l^{2} \rp)
  \sb=4\pi k^2 \l^{2}\sm~.
\eeq
Substituting this into Eq.~(\ref{eq:dm}), we obtain the equation of motion
for the magnetization:
\beq \label{eq:MxMypar}
  -i \o \,\sm=-4\pi g \bmo \tm \sm \lp[\a (1+l_d^2 k^2)-
  \frac{k^2\l^2}{1+k^2\l^2}\rp],
\eeq
with the dispersion relation
\beq \label{eq:disppar}
 \o=\pm \o_{fm}
  \left(1+l_d^2 k^2-{1\over \alpha} \frac{k^2\l^2}{1+k^2\l^2} \right)~,
\eeq
where $\o_{fm}=4\pi \a g M_0$ is
the frequency of the uniform FMR at zero {\it internal} magnetic
field for normal FM's.
In the limit $\l \rightarrow \infty$ the spin wave spectrum should
transform to that of an insulator. Neglecting
the spin
stiffness ($l_d \sim 0$) the latter can be presented as $\o= g(H_a +H)
$, where
$H_a= 4\pi M_0 \alpha$ is the anisotropy field \cite{akhi}. This agrees
with Eq.~
(\ref{eq:disppar}) (at finite $k$) bearing in mind that $H=-4\pi M_0$ in
our geometry (see Fig.
\ref{setup}). However, whatever large $\l$ could be, at
very small $k\ll 1/\l$ the dynamic induction is screened out, as  evident
from Eq. (\ref{eq:london}), so
at these scales the spin-wave spectra in superconducting and in insulating
ferromagnets are different.

The Maxwell equation $\partial  \sb/\partial t=-c \mathbf{\nabla}
\times \se $ and the London equation,
Eq. (\ref{eq:london}), relate the EM field inside the
sample to the magnetization:
\bear \label{eq:he}
  \sh&=&\sb-4\pi \sm=-\frac{4\pi}{1+k^2 \l^2}\sm~, \nonumber \\
  \se&=&\frac{\o}{k c} \sb \tm \bz = \frac{\o\l }{c}\frac{4 \pi k \l}{1+
   k^2 \l^2} \sm \tm \bz ~.
\eear

\begin{figure}
%\begin{center}
\includegraphics[width=0.3\textwidth]{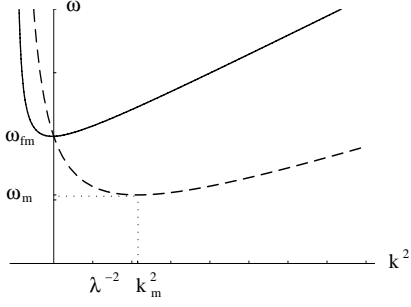}
%\end{center}
\caption{Form of the spectrum in two regimes: for $\sqrt{\a}\, l_d/\l=1.03$,
$\a=7$
(solid line); and
for $\sqrt{\a}\, l_d/\l=0.32$,  $\a=1.01$ (dashed line).}
\label{spectr}
\end{figure}

The two signs in Eq. (\ref{eq:disppar}) correspond to two senses of circular
polarization: $\sm^\ppm=m^\ppm(\bx \mp i\by)$.
Only  positively polarized waves $\sm^\pp$,
which correspond to the upper sign, can propagate inside
the sample, and in the following we shall
focus on this polarization. The form of the spectrum depends on  the ratio of
the
domain wall
thickness $l_d$ to the London
penetration depth $\l$, as shown in Fig.~\ref{spectr}. If
$\sqrt{\a}\,l_d>\l$, the minimal frequency (threshold for
spin-wave propagation) is $\o=\o_{fm}$ at $k=0$.
On the other hand, if $\sqrt{\a}l_d<\l$, then the minimal frequency is at a
finite wave vector,
$k_m= (1/\sqrt{\a}\,\l \,l_d -1/\l^2)^{1/2}$,
and  has a lower value,
%$\Oth=\a-(\sqrt{4\pi}-\g/\l)^2$.
$\o_m=\o_{fm}[1-(1/\sqrt{\a}-l_d/\l)^2]$. This dip in the spin-wave
spectrum was revealed by Ng and
Varma \cite{NV} for the spontaneous vortex phase. Waves with wave vectors
satisfying
$|k|<k_m$ have a  negative group velocity $d\o(k)/dk$, i.e., its sign is
opposite to that of the
phase velocity $\o(k)/k$. Thus in this regime
a positively polarized
incident EM wave with the frequency between $\o_m$ and
$\o_{fm}$ excites in the material two propagating waves, one with a positive
and another with a negative group velocity. This unusual property is unique to
SCFM's.

Now we turn to  calculation of the surface impedance, which specifies the
boundary conditions for the EM field outside the sample.
The dispersion  relation is a quadratic equation for $k^2$, so for a given
frequency
there should be two solutions for $k^2$: $k_1^2$
and $k_2^2$. These two solutions correspond to two modes generated by the
incident EM wave inside the sample. Thus the situation is
analogous to that
in {\em two mode electrodynamics},  the theory suggested for calculation of
the surface
impedance of a superconductor in the mixed state taking into account the
elastic degree
of freedom of the vortex array \cite{2M}. In the present case the second mode
(the second solution for $k^2$) originates from the spin degree of freedom.
Moreover,
in contrast to Ref. \cite{2M}, where both modes inside the sample were
evanescent
($k_{1,2}^2<0$), now at $\o>\o_{fm}$ one mode, and  at
$\o_m<\o<\o_{fm}$ both modes  are propagating.

In order to find a proper superposition of the two modes
inside the sample, $\sm^\ppm (z)=\sm_1^\ppm e^{i k_1 z} +\sm_2^\ppm e^{i k_2
z}$, we need
two boundary conditions. One of them is the usual continuity of the
EM field across the sample boundary.
%directly follows from the
%electrodynamics: The
%EM fields $\sh$ and
%$\se$ must be continuous at the sample boundary.
Polarization of the
excited spin wave is
defined by the polarization of the incident EM wave outside
the sample, and
the surface impedance tensor is
diagonal in the circular wave basis. In this basis $\sh=\hpm (z)\,(\bx\mp i
\by) $ and
$\se=\epm(z)\,(\bx\mp i\by) $, where $\hpm (z)=\hpm_1e^{i k_1 z}+\hpm_2e^{i
k_2 z}$
(and similarly for $\epm(z)$). Then the surface impedance is
\beq \label{eq:zpm}
  \zpm=\pm i \frac{\epm(0)}{\hpm(0)}~.
\eeq
The second boundary condition should be
imposed  on the magnetization. The simplest possible
boundary condition \cite{akhi} is $\pd \sm/\pd z=0$ at $z=0$,
which means
absence of spin currents through the sample surface. This gives
\beq
  k_1 \sm_1^\ppm+k_2 \sm_2^\ppm=0~.
\eeq
Together with Eq. (\ref{eq:he}) this yields the field
amplitudes at the surface:
\bear
  \epm(0)&=&
%\epm_1+\epm_2=
 \mp i\frac{4\pi \o \l}{c}\lp( \frac{k_1 \l}{1+k_1^2
  \l^2}-\frac{k_2 \l}{1+k_2^2 \l^2} \frac{k_1}{k_2}\rp) \mpm_1
\nonumber \\
  \hpm(0)&=&
%\hpm_1+\hpm_2=
-4\pi \lp(\frac{1}{1+k_1^2 \l^2}-
  \frac{1}{1+k_2^2 \l^2}\frac{k_1}{k_2}
  \rp)\mpm_1.
%\nonumber \\
%  \epm(0)&=&
%\epm_1+\epm_2=
% \mp i\frac{4\pi \o \l}{c}\lp( \frac{k_1 \l}{1+k_1^2
%  \l^2}-\frac{k_2 \l}{1+k_2^2 \l^2} \frac{k_1}{k_2}\rp) \mpm_1.
\eear
Substituting this into Eq. (\ref{eq:zpm}), we obtain after simplifications:
\beq \label{eq:imp}
 \zpm=-
\frac{\o \l^4}{c}\frac{k_1 k_2 (k_1+k_2)}{1+(k_1^2+k_2^2+k_1 k_2)\l^2}~.
\eeq
After  finding the two roots $k_1^2$ and $k_2^2$ of Eq.~(\ref{eq:disppar})
 we need to choose the signs of
$k_1=\pm \sqrt{k_1^2}$ and $k_2=\pm \sqrt{k_2^2}$. If the wave vectors have
imaginary parts, they should be positive in order to ensure attenuation
of the
wave $\propto e^{ik z-i \o t}$ inside the sample. But if $k$ is purely
real (propagating wave), the the sign  is fixed by the condition that
the energy must be carried away from the surface. One can check that the
energy flux for each mode
normal to the sample surface is proportional to the group velocity:
\beq
  P_z=-4\pi \a l_d^2\dot{\mathbf{m}}  {\partial \mathbf{m} \over \partial
z}  + {c\over 4\pi}
\mathbf{E}\times
  \mathbf{H}= {2 \pi \a \o\over \o_{fm}} |\mathbf{m}|^2\frac{d \o}{d k} ~.
      \label{Energ}\eeq
 Hence in order to carry energy away from the boundary, waves with a
 negative group velocity should
have wave vectors directed {\it toward} the boundary.
Bearing this in mind,
%it is
%straightforward to
we obtain from Eq. (\ref{eq:imp}) the general expression for
the surface impedance:
\beq
\zpm=\frac{\o \l}{c} \frac{\sqrt{Q(\a^{-1}-l_d^2 \l^{-2}+2 l_d
  \l^{-1}
  Q^{1/2}-Q)}}{\a^{-1}+ l_d \l^{-1} Q^{1/2}-Q }
\eeq
where $Q=1\mp\o/\ofm$, and the signs of the square roots are determined by
 the above requirements on the wave vectors 
(this is consistent with the condition $ \re\, \z>0$).

 Now we shall analyze the most
interesting
particular cases, when the frequency is close to thresholds at which spin-wave
propagation becomes possible. Two regimes should be considered, depending
on the value
of $l_d/\l$.

1. Stiff spin system: $\sqrt{\a}l_d> \l$. Then  the threshold  frequency
for spin-wave
propagation is $\o_{fm}$, and near the threshold $k_1 \ll |k_2|$:
\beq
k_1^2\simeq {\o -\o_{fm} \over \o_{fm}}{\a\over \a l_d^2 -\l^2}~,~~k_2^2
\simeq
{1\over \a l_d^2}-{1\over \l^2}~.
         \label{kk}     \eeq
Since $\sqrt{\a}l_d> \l$, the short-wavelength mode is
evanescent ($k_2^2<0$), while
the long-wavelength mode  is evanescent ($k_1^2<0$) 
at $\o <\o_{fm}$ and 
propagating ($k_1^2>0$) at $\o \gtrsim \o_{fm}$ . 
Finally the surface impedance is
\beq
  \z^\pp\simeq -
\frac{\o \l^4}{c}\frac{k_1 k_2^2}{1+k_2^2\l^2} = \frac{\sqrt{(\o
-\o_{fm})\o_{fm}}
}{c} \sqrt{\a(\a l_d^2 -\l^2)}~.
      \label{imp-soft}\eeq
Thus, $\z^\pp$ is purely
imaginary below the threshold and purely real above it, where it grows as
$\sqrt{\o -\o_{fm}}$
(actually, there is also nonzero $\im \, \z^\pp$ above
the
threshold, but
it is of higher order in small $\sqrt{\o -\o_{fm}}$).

2. Soft spin system: $\sqrt{\a}l_d< \l$. In this regime the
threshold
frequency has a lower  value $\o_m<\o_{fm}$. Near the threshold $|k_1|<k_m$ and
$|k_2|>k_m$ are
close to each other:
\beq
k_{1,2}^2\simeq k_m^2 \mp\sqrt{\o -\o_m \over \o_{fm}}{1\over \a^{1/4}
\l^{1/2}l_d^{3/2} }
~.
\eeq
Also, one should take into account that the mode with $k_1<k_m$ has a
negative
group velocity above the threshold, so the negative sign of $k_1$ should be
chosen.
Then $\z^\pp$ near the threshold is  given by
\beq
  \z^\pp\simeq
\frac{\o_m \l^4}{c}\frac{k_m^2 (k_2-|k_1|)}{1+\l^2k_m^2}=
\sqrt{\o -\o_m \over \o_{fm}} \frac{ \o_m  \l^2}{c\, l_d}\sqrt{ 1-
{\sqrt{\a}\,l_d\over
\l}}~.
\eeq
Again the surface impedance is real above the threshold
($\o >\o_m$)
and imaginary below it ($\o <\o_m$).

In addition, there is still a singularity at $\o=\o_{fm}$. We can use Eqs.
(\ref{kk}) and (\ref{imp-soft}), but now, since $\sqrt{\a}l_d< \l$, the
short-wavelength mode became propagating ($k_2^2>0$). The long-wavelength
mode is now
propagating ($k_1^2 >0$) at $\o<\o_{fm}$ and evanescent ($k_1^2 <0$) at
$\o>\o_{fm}$. Thus at $\o=\o_{fm}$ there is a transition from two propagating
modes to one.
 On the other hand, at $\o \gtrsim \o_{fm}$ Eq. (\ref{imp-soft})  yields
a purely
imaginary
impedance despite the presence of a propagating mode with real $k_2$.
In order to obtain the
real part of the surface impedance we need to go to  next order of the
expansion in the
small parameter $k_1/k_2$:
%\begin{eqnarray}
\beq
 \mbox{Re}\, \z^\pp(\o\gtrsim \o_{fm}) \simeq -
\frac{\o \l^4 k_1^2 k_2 }{c (1+k_2^2\l^2)^2 }
%\nonumber \\
=\frac{(\o
-\o_{fm}) \a^{5/2} l_d^3 }{  \l\, c \sqrt{\l^2-\a l_d^2}}.
\eeq
%\end{eqnarray}
Hence at $\o=\o_{fm}$, $\mbox{Re}\z^\pp$ changes its behavior from $\sim
\sqrt{\o_{fm}-\o}
$
below $\o_{fm}$ to $\sim (\o-\o_{fm}) $ above $\o_{fm}$.

Our analysis can be easily generalized to take into account dissipation
due to normal currents. For this, a normal current should be added to the
SC current in the Maxwell equations. As a result, a
renormalized complex $\lt$ should be
used instead of the London penetration depth: $ \lt^{-2}=\l^{-2}-2i \d^{-2}$,
where $\d=c/\sqrt{2 \pi \s_n \o}$ is the skin depth, and $\s_n$ is the normal
conductivity. Expanding with respect to the small ratio $\l^2/\d^2$, we
obtain instead of  Eq. (\ref{imp-soft}):
\beq
  \z^\pp \simeq \frac{\sqrt{\a(\o -\o_{fm})\o_{fm}}
}{c}\lp[ \sqrt{(\a l_d^2 -\l^2)}-i  {\l^4 \over \d^{2}}\sqrt{1 \over \a
l_d^2 -\l^2}\rp]
       \eeq
In the opposite limit  $\l \rightarrow \infty$,
$ \lt^{2}=i \d^{2}/2$, and we
obtain the surface impedance for normal FM metals \cite{ament}. However, unlike
in SCFM's,
in
FM metals the  dissipation is an essential part of spin
dynamics, which can not be assumed small.
%cannot propagate  with so low attenuation as in superconducting
%ferromagnets.

Note that at
$\o=\o_{fm}$ [see Eq. (\ref{imp-soft})], $\z^\pp$ vanishes even
after renormalization of the
penetration depth, since at uniform FMR no currents are generated, so the
dissipation due to
normal currents is ineffective.
% because no currents are generated.
Then, other sources of dissipation become important, in particular, the
transverse spin
relaxation, which is well known from studies of normal
ferromagnets \cite{akhi}.

 Our phenomenological approach is valid for
any material with coexisting SC and FM order parameters, independent
of their microscopic origins.
Moreover, we even believe that  it is not essential whether FM
originates from electron spins or
from the orbital moment of Cooper pairs, which characterizes the $p$-wave
pairing in some materials
\cite{mac}. In the latter case one cannot call the modes which were
analyzed here, ``spin waves'': they
should be called  ``orbital waves'' as in $^3$He physics. It is known from
$^3$He studies that the dynamics
of the orbital moment is similar to spin dynamics \cite{He}. In the
case of electron pairing, when both the spin
and  orbital  motion contribute to the magnetic moment, one can consider
the dynamics
of the total  magnetic moment. Consequently, the corresponding excitations
will be ``magnetization waves''.
%unite two types
%of waves under the name
%``magnetization waves''.
In the case of coexistence of SC
with  orbital FM
the term ``ferromagnetism'' is sometimes avoided, while the corresponding
category of materials
is commonly called
``superconductors with broken
time-reversal symmetry''.  However, independently of this semantic variance,
our analysis is relevant for these
materials.

In conclusion, we have demonstrated that propagating spin waves are
possible in superconducting
ferromagnets in the Meissner state, similar to those in insulating
ferromagnets. In contrast to the latter,
 in SCFM's with low spin stiffness there
is a frequency interval, in
which there are {\em two} propagating spin modes of the same circular
polarization, one of them having a
negative group velocity, i.e., with the direction opposite to that of the
phase velocity. We solved the
boundary problem and
calculated the surface impedance for the case of the easy
magnetization  axis normal to
the surface of the sample. Revealed singularities of the dependence of the
surface impedance on the frequency
provide experimentalists with a tool for probing spin-wave dynamics by
microwave measurements of
materials with coexisting FM and SC.

\begin{acknowledgments}
This work was supported by the grant of the Israel
Academy of Sciences and Humanities.
\end{acknowledgments}

\end{document}